\documentclass[draft,jgrga]{AGUTeX}

 \usepackage[dvips]{graphicx}
  \setkeys{Gin}{draft=false}
\authorrunninghead{SHARE ET AL.}

\titlerunninghead{Unfounded Solar Neutron Claim}

\begin{document}

%% ------------------------------------------------------------------------ %%
%
%  TITLE
%
%% ------------------------------------------------------------------------ %%

%\title{Claimed {\textit {MESSENGER}} detection of solar neutrons on 2011 June 4 is unfounded}
\title{Comment on ``Detection and characterization of 0.5--8 MeV neutrons near Mercury: Evidence for a solar origin"}
%
% e.g., \title{Terrestrial ring current:
% Origin, formation, and decay $\alpha\beta\Gamma\Delta$}
%

%% ------------------------------------------------------------------------ %%
%
%  AUTHORS AND AFFILIATIONS
%
%% ------------------------------------------------------------------------ %%

%Use \author{\altaffilmark{}} and \altaffiltext{}

% \altaffilmark will produce footnote;
% matching \altaffiltext will appear at bottom of page.

% \authors{A. B. Smith,\altaffilmark{1}
% Eric Brown,\altaffilmark{1,2} Rick Williams,\altaffilmark{3}
% John B. McDougall\altaffilmark{4}, and S. Visconti\altaffilmark{5}}

%\altaffiltext{1}{Department of Hydrology and Water Resources,
%University of Arizona, Tucson, Arizona, USA.}

%\altaffiltext{2}{Department of Geography, Ohio State University,
%Columbus, Ohio, USA.}

%\altaffiltext{3}{Department of Space Sciences, University of
%Michigan, Ann Arbor, Michigan, USA.}

%\altaffiltext{4}{Division of Hydrologic Sciences, Desert Research
%Institute, Reno, Nevada, USA.}

%\altaffiltext{5}{Dipartimento di Idraulica, Trasporti ed
%Infrastrutture Civili, Politecnico di Torino, Turin, Italy.}

\authors{Gerald H. Share\altaffilmark{1}, Ronald J. Murphy\altaffilmark{2},  Allan J. Tylka\altaffilmark{3},  Brian R. Dennis \altaffilmark{4}, and
James M. Ryan\altaffilmark{5} }

\altaffiltext{1}{Astronomy Department, University of Maryland, College Park, MD 20742, USA. (share@astro.umd.edu)}

\altaffiltext{2}{Space Science Division, Naval Research Laboratory,  Washington, DC 20375, USA. (ronald.murphy@nrl.navy.mil) }

\altaffiltext{3}{Code 672, NASA Goddard Spaceflight Center,  Greenbelt, MD  20771., USA. (allan.j.tylka@nasa.gov) }

\altaffiltext{4}{Code 671, NASA Goddard Spaceflight Center,  Greenbelt, MD  20771., USA. (brian.r.dennis@nasa.gov) }

\altaffiltext{5}{Space Science Center, University of New Hampshire, Durham, NH 03824, USA.  (james.ryan@unh.edu)}

%% ------------------------------------------------------------------------ %%
%
%  ABSTRACT
%
%% ------------------------------------------------------------------------ %%

% >> Do NOT include any \begin...\end commands within
% >> the body of the abstract.

\begin{abstract}

We argue that the hour-long neutron transient detected by the MErcury Surface, Space ENvironment, GEochemistry, and Ranging ({\it MESSENGER}) Neutron Spectrometer beginning at 15:45 UT on 2011 June 4 is due to secondary neutrons from energetic protons interacting in the spacecraft.  The protons were probably accelerated by a shock that passed the spacecraft about thirty minutes earlier.  We reach this conclusion after a study of data from the {\it MESSENGER} neutron spectrometer, gamma-ray spectrometer, X-ray Spectrometer, and Energetic Particle Spectrometer, and from the particle spectrometers on {\it STEREO A}.  Our conclusion differs markedly from that given by \citet{lawr14} who claimed that there is ``strong evidence" that the neutrons were produced by the interaction of ions in the solar atmosphere.  We identify significant faults with the authors' arguments that led them to that conclusion. 

%associated with the passage of a shock and CME. 

\end{abstract}

%% ------------------------------------------------------------------------ %%
%
%  BEGIN ARTICLE
%
%% ------------------------------------------------------------------------ %%

% The body of the article must start with a \begin{article} command
%
% \end{article} must follow the references section, before the figures
%  and tables.

\begin{article}

%% ------------------------------------------------------------------------ %%
%
%  TEXT
%
%% ------------------------------------------------------------------------ %%

\section{Introduction}
Neutrons resulting from interactions of flare-accelerated ions in the solar atmosphere and escaping from the Sun carry important information about the ions responsible for their production \citep{murp12}. Because low-energy (1-10 MeV) solar neutrons decay before they reach Earth, they can only be measured in the inner heliosphere.  The MErcury Surface, Space ENvironment, GEochemistry, and Ranging ({\it MESSENGER}) spacecraft contains a neutron spectrometer  that is able to unambiguously detect such low-energy neutrons \citep{feld10}.  In a recent paper, \citet{lawr14} claimed ``strong evidence" that neutrons detected with this spectrometer during an hour-long transient beginning at 15:45 UT on 2011 June 4, when {\it MESSENGER} was in orbit around Mercury at a distance of 0.33 AU, originated at the Sun.  This transient in the valid neutron event rate is plotted in Figure \ref{particles}a (adapted from Figure 4 of \citet{lawr14}). 
 
The \citet{lawr14} main argument that the neutrons detected during the transient were produced at the Sun, and not locally by interaction of energetic ions with the spacecraft, is that their calculations show that the number of ions at {\it MESSENGER} were insufficient to produce such a large increase in secondary neutrons.  This conclusion is based on the upper limit they obtained on the number of $>$45 MeV protons at {\it MESSENGER} at the time of the transient and assumptions about the proton spectrum and angular distribution.  They further argued that the gamma-ray spectrum observed during the transient is consistent with neutrons interacting in the spacecraft, but not with protons.  Notably, they admitted that they ``cannot rule out the presence of lower-energy ions ($<$30 MeV) that can produce local neutrons." They also admitted that if the neutrons originated at the Sun, their measured flux at MESSENGER during the transient requires an ``unexpectedly large" number of low-energy ions interacting at the Sun.

In this Comment we question the evidence cited by \citet{lawr14} supporting their claim that the neutrons producing the observed transient came from the Sun.  Specifically, we use data presented in the \citet{lawr14} paper, along with complementary observations, to demonstrate that there was a sufficient flux of $>$1 MeV protons interacting in the spacecraft at the time of the transient to produce the number of neutrons detected.   Further, the authors' claim that only neutrons can produce the gamma-ray line spectrum during the transient is contrary to measured neutron and proton cross sections for nuclear excitations.  Thus, the authors' arguments supporting a solar origin of the detected neutrons either have serious flaws or, if viewed properly, actually support local production of the neutrons by interactions of MeV to tens of MeV nucleon$^{-1}$ ions with the spacecraft. These ions were probably accelerated by a shock, associated with a coronal mass ejection (CME), that passed Mercury 30 min prior to the onset of the transient \citep{lari13}.      

In $\S$2 we show that charged-particle detectors on {\it MESSENGER} and {\it STEREO A} responded to similar SEP particle populations from a solar eruption commencing at 07:06 UT \citep{nitt13} when the two spacecraft were on nearly the same magnetic field line connected to the flaring region on the Sun \citep{lari13}.  With confidence in the {\it MESSENGER} particle measurements, we  then demonstrate that the transient was due to secondary neutrons produced by the ion interactions in the spacecraft.  In $\S$3 we point out flaws in the authors' arguments supporting their claim that the transient was from neutrons originating at the Sun.  In $\S$4 we summarize our arguments in favor of neutrons produced as secondaries in the spacecraft as the source of the transient and against neutrons produced at the Sun.  If solar neutrons had been present at {\it MESSENGER} with the claimed fluxes, explaining their origin would be challenging given that the only high-energy activity on the Sun was the flare and eruptive event starting at 07:06 UT which occurred 9 hours prior to the transient.  We also discuss the prospects for future detection of solar neutrons with {\it MESSENGER}. 

\section{Energetic ions interacting in {\textit {MESSENGER}} produced the neutron transient at 16 UT }

In this section we present evidence that MeV to tens of MeV nucleon$^{-1}$ ions at {\it MESSENGER}, probably accelerated by a passing shock from the CME associated with an M5.2--X1.6 solar flare (N15W140) at 07:06 UT \citep{nitt13}, produced secondary neutrons in the spacecraft beginning at 15:45 UT.  We use data obtained with the same instruments used by \citet{lawr14}.  These include the {\it MESSENGER} neutron spectrometer (NS) \citep{gold07}, the Gamma Ray Spectrometer (GRS) \citep{gold07},  the X-Ray Spectrometer (XRS) \citep{schl07}, and the Energetic Particle Spectrometer (EPS) \citep{andr07} and the {\it STEREO A} In-situ Measurements of Particles and CME Transients (IMPACT) high-energy telescope (HET)\citep{luhm08}.  

Figure \ref{particles} summarizes the relevant data rates and fluxes observed by the instruments on 2011 June 4 from 05 to 20 UT.  NS consists of two lithium-glass (LG1 and LG2) scintillators and one borated plastic (BP) scintillator that are each sensitive to $>$15 MeV protons and $>$1 MeV electrons \citep{lawr14}.  \citet{feld10} provided the details about how these detectors measure neutrons.  As discussed above, in panel a) we plot the valid neutron count rates from NS that shows the transient beginning at 15:45 UT and an earlier peak near 12 UT, during a periapsis transit of Mercury, due to neutrons produced by cosmic-ray interactions in the planet's surface. In panel b) we plot the count rates from ionizing particles in the LG and BP scintillators (singles rates) as given in Figure 6 of \citet{lawr14}.  In panel c) we plot what \citet{lari13} identified as the 71--112 keV electron fluxes from EPS.  We note that while EPS was designed to detect electrons from 0.025--1.0 MeV, its nominal electron-energy channels also respond to $>$110 keV protons and $>$1 MeV electrons \citep{andr07}.  In their Figure 19, \cite{lawr14} plotted the unfiltered XRS count rates in support of their claim.  XRS is comprised of three gas proportional counters for studying fluorescent X-ray emission from Mercury.   To improve its sensitivity to fluorescent X-rays, pulse-shape discrimination is used to identify traversing charged particles which produce pulse rise-times that are longer, due to their longer path lengths, than those from X-rays.  We plot the count rate from this charged-particle channel in Figure \ref{particles}c).  In Figure \ref{particles}d) we plot 5-min averaged fluxes of 0.7--2.8 MeV electrons, and 13--26, 26--40, and 40--100 MeV protons observed with IMPACT on {\it STEREO A} at 1 AU. 

\subsection{Similar solar energetic particle environments at {\textit {MESSENGER}} and {\textit {STEREO A}} prior to the neutron transient }

The solar energetic particle (SEP) environments at {\it MESSENGER} and {\it STEREO A} on June 4 were both dynamic prior to, during, and after the time of the neutron transient at 16 UT.  Mercury was aligned magnetically with an active region behind the west limb of the Sun as viewed from Earth \citep{lari13}.  This region produced two solar eruptive events (SEEs) starting at 07:06 and 21:51 UT \citep{nitt13} with associated CMEs having speeds of about 1400 and 2400 km s$^{-1}$, respectively.  

{\it STEREO A}, at 1 AU from the Sun, was situated close to the same magnetic field line as {\it MESSENGER} (the heliolongitudes of the footpoints of the interplanetary magnetic field lines connecting {\it MESSENGER} and {\it STEREO A} with the Sun were within 3$^{\circ}$ of one another \citep{lari13a}).   Relativistic 0.7--2.8 MeV electrons from the first SEE began arriving at {\it STEREO A} at about 07:30 UT and lower velocity 13--100 MeV protons began arriving after 08:15 UT (Figure \ref{particles}d).  The electron flux plateaued near 09:00 UT and fell slowly until about 16:15 UT at which time it again slowly increased.  The {\it STEREO A} proton flux reached a peak near 11:30 UT, dropped by about 30\% at energies below 40 MeV, and plateaued until about 16:15 UT when it again increased.   Thus, there is clear evidence for the presence of energetic electrons and protons at {\it STEREO A} following the 07:06 UT SEE and evidence for significant flux variability in both particle species.  Below we show that  there was a similar energetic particle environment at {\it MESSENGER}, as expected, because it was located close to the field line connecting {\it STEREO A} and the Sun.

The LG, and BP singles rates (Figure \ref{particles}b) and XRS particle-channel rates (Figure \ref{particles}c) all increased after about 07:15 UT, with the arrival of SEPs from the 07:06 UT SEE.  Relativistic electrons should have arrived at {\it MESSENGER} by this time because they were first detected at 1 AU by {\it STEREO A} at 07:30 UT.  To be consistent with the proton increase at {\it STEREO A} at 08:15 UT, protons with energies $>$15 MeV should have started arriving at {\it MESSENGER} at about 07:30 UT because such protons would take about 45 min to travel between Mercury and 1 AU.   The LG and BP singles and XRS particle rates show some structure at the time of the Mercury periapsis transit at 12 UT and additional variability after that time. (The spikes in the XRS during the transit are due to the automatic lowering of its high voltages to protect the detectors from high count rates.)  

We can compare the charged particle intensities at {\it STEREO A} and {\it MESSENGER} to determine whether the SEP environments at the two locations were consistent with one another.  Between 12 -- 14 UT {\it STEREO A} observed a 13 -- 100 MeV proton flux of about 5 cm$^{-2}$ s$^{-1}$ sr$^{-1}$.    Because the two spacecraft were aligned close to the same magnetic field line we do not expect a significant longitudinal flux difference.   \citet{lari13a} reported about a factor of five increase in SEP flux at Mercury compared with that observed with {\it STEREO A} following the 07:06 UT SEE, due to its closer proximity to the Sun.  Assuming a 1-sr acceptance angle for the {\it MESSENGER} detectors, we estimate that the 100-cm$^{2}$ LG2 detector would have observed a proton rate of about 2500 counts s$^{-1}$ (5 cm$^{-2}$ s$^{-1}$ sr$^{-1}$ $\times$ 1 sr  $\times$ 100 cm$^{2}$ $\times$ 5).   The measured LG2 singles rate was close to 1000 counts s$^{-1}$, consistent with this estimate given the uncertainties in the anisotropy of the SEP protons and the spacecraft shielding.  We note that the frontal areas of the LG detectors are oriented orthogonal to the {\it MESSENGER}/Sun direction and that the SEP particles are expected to primarily come from the hemisphere in the solar direction.  Thus, the estimated LG2 singles count rate could be lower than that calculated above because of the reduced projected detector areas and also because of the spacecraft material through which the protons would need to pass in order to reach NS. The 0.7 -- 2.8 MeV electron flux at {\it STEREO A} between 12 -- 14 UT was about 0.4 cm$^{-2}$ s$^{-1}$ sr$^{-1}$.  Following the estimate made for protons, this flux would have yielded a rate of about 200 counts s$^{-1}$ in LG2, about 10\% of the rate from protons.  Thus, the LG, BP, and XRS detectors were probably responding mostly to protons during that time.

Based on these timing and flux comparisons, the SEP environments at {\it MESSENGER} and {\it STEREO A} were consistent with one another prior to the neutron transient at 16 UT.    This contrasts with the assertion by \citet{lawr14} that the two spacecraft ``were in markedly different energetic charged-particle environments," although they did admit that this would be ``an unusual situation for pairs of locations on similar field lines \citep{lari13a}."  

\subsection{Evidence that energetic ions interacted in {\textit {MESSENGER}} to produce the neutron transient}

An interplanetary shock from the 07:06 UT SEE passed {\it MESSENGER} at about 15:10 UT followed by its CME at about 17:45 UT \citep{lari13}.  The LG and BP singles rates (Figure \ref{particles}b) and the EPS electron-channel fluxes (Figure \ref{particles}c)) all show striking peaks beginning about 15:45 UT, about 35 min after passage of the shock, that are remarkably similar to the peak in the neutron count rate plotted in Figure \ref{particles}a.  The XRS particle rate (Figure \ref{particles}c) also shows an intense peak near 16 UT but with more complex structure in part due to the fact that the high voltage was lowered beginning at 16 UT to protect the counters.  Because of this high-voltage reduction we do not know what the peak XRS rate was, but it clearly exceeded 1000 counts s$^{-1}$.  Because XRS has an area of 30 cm$^{2}$, its peak rate appears to be consistent with the $\sim$ 9,000 counts s$^{-1}$ in the 100 cm$^{2}$ LG2 detector, assuming that they were both responding to the same particles and were comparably shielded.  Thus, there was a large flux of charged particles present at {\it MESSENGER} following the same time profile as the neutron transient.

We now present evidence that a significant fraction of the charged particles causing the high LG- and BP-singles, XRS, and EPS rates during the neutron transient were ions and not electrons.  In the top panel of their Figure 10 (our Figure \ref{gamspec}) \cite{lawr14} plotted the energy-loss spectrum recorded by the {\it MESSENGER} gamma-ray spectrometer during the 1-hr neutron transient at 16 UT compared to an earlier background spectrum produced by cosmic rays.  The gamma-ray flux recorded at the time of the transient was a factor of 5-10 above background at all energies from 500 keV to 4.5 MeV.   The transient gamma-ray spectrum shows the characteristic de-excitation lines produced by inelastic and spallation interactions of protons and/or neutrons with spacecraft materials.  In contrast, the spectrum did not show the steeper bremsstrahlung continuum that electrons would produce.  Therefore, a significant fraction of the charged particles producing the high rates in the detectors and interacting in the spacecraft during the neutron transient had to be ions.

We next determine whether the peak neutron count rate is consistent with the LG singles rates if the transient were due to secondary neutrons produced by ion interactions in the satellite.  The maximum effective area of the LG detectors for low-energy protons must be close to its 100 cm$^{2}$ geometric area.  In contrast, the NS effective area for detecting 1-10 MeV neutrons is not likely to be higher than about 10 cm$^{2}$ (we could not find any information on the absolute efficiency of the detector \citep{gold07,feld10,lawr14}).  \citet{feld10} estimated that it would require about 100 20-MeV protons passing through the spacecraft to produce one $>$1 MeV neutron. Because the peak LG2 rate was 9000 counts s$^{-1}$, we estimate that the NS neutron rate would be about 9 counts s$^{-1}$ (9000 s$^{-1} \times 0.01 \times 0.1$).  This is consistent with the 15 counts s$^{-1}$ rate observed. 

\citet{lawr14} noted that relative differences between the rates in the LG1 and LG2 detectors could be due to their different viewing geometries.  The detectors are on either side of BP, which is thick enough to stop a 125 MeV  proton.  Thus, we can obtain information on the isotropy of lower-energy ions interacting in {\it MESSENGER} by comparing the LG2/LG1 singles rate ratios.   The LG2/LG1 rate ratio was about unity up until 12 UT (Figure \ref{particles}b), suggesting a relatively isotropic angular distribution for protons reaching NS, at least orthogonal to the Mercury-Sun line.  The LG2/LG1 ratio rose to a peak of 5.5 near 16 UT and then decreased suggesting that the angular distribution of the protons reaching NS was asymmetric orthogonal to the Sun-Mercury direction at the time of the neutron transient. This suggests that the angular distribution of the incident protons during the transient, and following the passage of the shock, was significantly different than at other times. 

Based on the above arguments, we conclude that the high rates observed in the charged-particle detectors on {\it MESSENGER} at the time of the neutron transient were primarily due to energetic protons (and alpha particles), likely accelerated by the passing shock related to the solar eruption at 07:06 UT.  We, therefore, also conclude that the neutrons observed by NS during the transient were secondaries produced by the interactions of these energetic ions in the spacecraft .

\section{Unsupportable arguments for solar neutrons on 2011 June 4}

We now address the \citet{lawr14} arguments for concluding that the neutrons observed with NS during the transient peaking at 16:15 UT were from the Sun. We show that these arguments are either not supported by the data or have alternative explanations.  The authors' evidence against a secondary origin for the neutrons from ion interactions in the spacecraft is based on lack of a time coincident $>$45 MeV proton rate increase and calculations that they claimed rule out the presence of enough low-energy protons to produce the neutrons.   The calculations are complicated and are based on various assumptions that we discuss in $\S$3.3.    \citet{lawr14} also argued that the gamma-ray spectrum detected by GRS during the neutron transient was produced by neutrons and not by protons.  We dispute this contention in $\S$3.2. We first discuss the high rates in the {\it MESSENGER} LG, BP, EPS, and XRS detectors that we have shown to be due to protons, but whose origin according to \citet{lawr14} is not understood.

\subsection{Unexplained origin of the high charged-particle rates if the neutrons came from the Sun }

\citet{lawr14} did not offer a viable explanation for the high singles rates in the LG and BP detectors shown in Figure \ref{particles}b, and in the EPS and XRS detectors shown in Figure \ref{particles}c during the neutron transient.  If the transient were due to solar neutrons, then the high singles rates must be due to secondaries produced by neutron interactions in the spacecraft.  We argue here that these high rates could not be due to secondary protons produced by solar neutron interactions. First, because the neutron energies during the transient barely exceeded 7 MeV, as shown in Figure 12 of \citet{lawr14}, neutron interactions could not possibly have produced protons with energies $>$15 MeV that are required for them to be detected in the LG and BP detectors.  Second,  the peak neutron rate was too small to have produced the rates of charged particles detected in the LG detectors.   For an assumed neutron detector effective area of 10 cm$^2$, the peak neutron flux would have been 1.5 cm$^{-2}$ s$^{-1}$.  For a peak charged-particle flux of 90 cm$^{-2}$ s$^{-1}$ (9000 cts s$^{-1}$/100 cm$^2$) in the LG2 detector, each neutron would have had to interact in {\it MESSENGER} to produce 60 secondary protons. In striking contrast, \citet{feld10} estimated that it would take 100 proton interactions in the spacecraft to produce one secondary neutron. As the nuclear interaction cross sections for protons and neutrons are comparable, it should also take about 100 neutrons to produce each secondary proton. Thus, it is inconceivable that solar neutrons could have produced such a high charged-particle count rate in the LG detectors from secondary protons.  
 
The caption of Figure 19 in \citet{lawr14} suggests that the increase seen in the unfiltered XRS detector rates near 16 UT is due to secondary gamma-rays produced by the solar neutrons.   Following our arguments above, each solar neutron would then have had to produce enough 1--10 MeV gamma rays in the spacecraft in order to account for 60 secondary $>$1 MeV electrons in LG2. As a 2 MeV gamma ray has an $\sim$ 5\% probability of interacting in the 0.4 cm LG detector, this would require that each neutron produce about 1000 secondary gamma rays in {\it MESSENGER}.  Thus, it is also inconceivable that solar neutrons could have produced such a high LG count rate from secondary gamma rays. 

In contrast to what they wrote in the caption for Figure 19, the authors ruled out a secondary gamma-ray origin for the peak in the LG and BP singles rates because there were no comparably high increases in these rates at 12 UT during the close-approach to Mercury when the neutron rate was almost as high as it was during the transient at 16 UT (see Figure \ref{particles}a).   Having ruled out a gamma-ray origin for the increases in the singles, EPS, and XRS rates during the transient, the authors admitted that ``the large singles count rates allow for the presence of $<$45 MeV protons and/or $<$10 MeV electrons during the neutron event."  Taking everything into account, we cannot understand how \citet{lawr14} then concluded that there is ``strong evidence" that the transient was due to neutrons produced at the Sun.

\subsection{Errors in interpreting the gamma-ray spectrum and its origin}

Figure \ref{gamspec} reproduces the gamma-ray spectrum during the transient shown in Figure 10 of \citet{lawr14}. The authors attributed all the de-excitation lines, such as $^{12}$C, $^{16}$O, and $^{27}$Al, to inelastic neutron interactions and stated that lines from spallation reactions can only be produced by protons.  This is not correct.  In fact, the neutron and proton cross sections for producing all these lines are similar.  For example, the cross sections for producing the de-excitation line at 4.43 MeV from inelastic neutron and proton reactions with $^{12}$C both reach maximum values of ~350 mb near 10 MeV (see Figure 1 in \citet{murp11} and Figure 4 in \citet{kozl02}).   In contrast to what the authors stated, spallation lines such as the 1.634 MeV $^{20}$Ne line can also be produced by both neutron and proton interactions with Mg in the spacecraft.  For these reasons, none of their arguments about the gamma-ray spectrum used to rule out the presence of one to tens of MeV protons during the time of the neutron transient are valid. Protons will produce an almost identical gamma-ray spectrum so there is no need to require the presence of a solar neutron flux at the time of the transient. Above we ruled out the presence of such a solar neutron flux because it could not produce the high charged-particle rates observed in the LG, XRS, and EPS detectors during the transient.

It is surprising that the authors did not discuss whether a similar de-excitation line gamma-ray spectrum was observed when {\it MESSENGER} was close to Mercury at 12 UT when comparable neutron count rates were present.  We note that the data points at 12 UT are missing in their Figure 10b showing the time histories of various de-excitation lines.  As noted above in $\S$3.1, \citet{lawr14} ruled out secondary gamma rays from neutron interactions as the origin of the increases observed in the LG, BP, EPS, and XRS detectors at the time of the neutron transient at 16 UT because such large increases were in fact not observed at Mercury periapsis near 12 UT.   We do not understand why the authors did not apply the same argument for the GRS spectrum during the transient.

There is, therefore, no evidence that the intense gamma-ray spectrum observed around 16 UT was produced by solar neutrons interacting with the spacecraft, as \cite{lawr14} contended.  Rather it seems almost certain that the spectrum was generated by ions, likely associated with the passing CME shock, interacting with the spacecraft.

\subsection{Authors' arguments about the proton spectrum at {\textit MESSENGER}}

We next address the \citet{lawr14} arguments that their study of $>$45 MeV protons rules out the presence of sufficient lower energy protons at  {\it MESSENGER} to produce the neutron transient.   They plotted these  LG1-BP and LG2-BP double-coincidence rates in Figure 7 of their paper, along with the LG1-BP-LG2 triple coincidence rates that are sensitive to $>$125 MeV protons. There is significant background in these channels due to the predominance of Galactic cosmic-rays at these energies.  \citet{lawr14} noted increases in the double coincidence rates at about 08 UT; these are similar to the increases in the singles and XRS rates due to the arrival of SEPs from the 07:06 UT SEE that we discussed in $\S$2.1.  Thus, there was a significant flux of SEP ions up to energies in excess of 45 MeV present at {\it MESSENGER} before the transient.  We also note that fluxes in gamma-ray lines observed with GRS appear to rise after about 08:00 UT (see Figure 10b) of their paper).  These lines are produced by interactions of ions in the spacecraft, providing further evidence for SEPs at {\it MESSENGER} prior to the transient.  \citet{lawr14} pointed out significant decreases in the fluxes of $>$45 and $>$125 MeV protons after about 15 UT, which they did not explain. These are probably due to a Forbush decrease in the cosmic-ray intensity from the passage of the same shock that we believe produced the lower-energy charged-particles responsible for the neutron transient beginning at 15:45 UT. 

At the time of the neutron transient \citet{lawr14} could not find evidence for a peak in the double coincidence rate sensitive to $>$45 MeV protons.   They then used the upper limit on $>$45 MeV proton flux and assumptions about the charged-particle spectrum, $\alpha$/p ratio, and angular distribution to rule out charged particles as the source of the neutron transient.   Similar calculations were employed by \citet{feld10} to justify their claimed detection of solar neutrons by {\it MESSENGER} on 2007 December 31, but were questioned by \citet{shar11} because their calculations did not extend to proton energies below 30 MeV and did not include the presence of $^{13}$C in the spacecraft carbon fiber structure.  \cite{lawr14} stated that these corrections have now been made but did not make clear how low in proton energy the calculations were performed.  The calculations must be made down to 1 MeV nucleon$^{-1}$ because alpha particle interactions are an important source of neutrons at MeV energies.

In any case, \citet{lawr14} argued in Figure 9 of their paper that the observed transient neutron count rate exceeded the calculated rate due to proton interactions by a factor of $\sim$750.  However, the calculated value was obtained for a hard SEP spectrum typically produced by shocks starting within a few R$_{\odot}$ of the Sun and not for a locally-produced shock spectrum that is usually significantly softer \citep{desa99,ream12}.  For such a softer shock-produced spectrum, \citet{lawr14} found that the observed/calculated ratio drops to $\sim$ 10.  However, they assumed an isotropic particle distribution. The authors then considered the case of an anisotropic distribution assuming that the particles came from the direction of the Sun and therefore passed through a significant amount of material before reaching  NS.  For this anisotropic distribution they concluded that the observed/calculated neutron count ratio would then increase by a factor of 5 to 10.  We note, however, that the receding shock that passed {\it MESSENGER} would produce energetic particles from the anti-solar hemisphere.  In this case, the observed/calculated ratio would likely {\it decrease} by a factor of 5 to 10. With this correction, the observed and calculated neutron count rates would be comparable. Therefore, their argument for ruling out $<$45 MeV protons as the source of the neutron transient is not compelling.  More importantly, we showed in $\S$2 that such low-energy protons are indeed present at Mercury during the neutron transient, based on the high charged-particle rates observed in the LG and BP scintillators, the XRS charged-particle channel, and the EPS spectrometer  

\section{Discussion}

We provide conclusive evidence that the neutron transient observed by {\it MESSENGER} for an hour beginning about 15:45 UT on 2011 June 4 was not due to neutrons from the Sun, as claimed by \citet{lawr14}, but to neutrons produced by ions interacting in the spacecraft following the passage of a CME shock.  We also demonstrate that critical arguments made by \citet{lawr14} supporting the solar origin of the neutrons either are unsubstantiated or have serious flaws.  

From the lack of a time coincident flux of protons $>$45 MeV during the transient, \cite{lawr14} argued that there were insufficient numbers of protons producing secondary neutrons to account for the observed neutron count rate.  Using realistic choices for the proton spectrum and angular distribution, we find that the rates of observed and estimated neutrons are consistent with one another. \citet{lawr14} did not explain the remarkable time-coincidence between the neutron transient and $>$15 MeV proton rates but did admit that these protons could produce local neutrons.  We estimate that the observed $>$15 MeV proton count rate is indeed sufficient to produce the observed neutron count rate.   If the transient had been produced by neutrons from the Sun, as the authors suggest, then each one of these 1--10 MeV neutrons would have had to produce about 60 $>$15 MeV protons to account for the observed charged particle rates.  

Another key argument made by \citet{lawr14} is that the measured gamma-ray spectrum during the neutron transit is consistent with production by neutrons and not by protons.  We are baffled by this contention because the neutron and proton cross sections for producing all the lines in the spectrum are nearly identical. 

\citet{lawr14} is the second paper in which the the {\it MESSENGER} NS team claimed the detection of solar neutrons during an ongoing solar energetic particle event.  \citet{feld10} made that claim for the 2007 December 31 limb flare using data from {\it MESSENGER} when it was at 0.48 AU and presented calculations purporting to show that secondary neutrons from the SEPs could not account for the measured neutron count rate.  \citet{shar11} studied the evidence for the detection of solar neutrons and concluded that \citet{feld10} had underestimated the number of secondary neutrons produced by solar protons and alpha particles interacting in the spacecraft and found that there was no basis for their claim.  \citet{lawr14} have recalculated the number of secondary neutrons produced using ``appropriate low-energy cross sections" and conclude ``that the measured neutrons are still more than a factor of 2 larger than the upper limit on the production by SEP ions."  As discussed in $\S$3.3 there are significant uncertainties in these calculations due to various assumptions that they used including the assumed proton spectrum, angular distribution, and $\alpha$/p ratio.  Thus, this factor of 2 difference is not sufficient to require the presence of neutrons produced at the Sun. 

It is difficult to understand how the neutrons detected in the transients on 2007 December 31 and 2011 June 4 could have been produced if they had indeed been solar.  The timing of both transients indicates that the neutrons would have needed to be produced at the Sun hours after any observed flaring-energy activity. Such flaring activity is typically needed to accelerate ions to energies $\geq$1 MeV nucleon$^{-1}$ necessary to produce the neutrons. This led \citet{lawr14} to invoke some gradual nuclear interaction process with a low-energy threshold in the solar corona or chromosphere that is not readily explained physically,  especially because of the extraordinarily high number of protons ($10^{34} >$30 MeV) at the Sun required to account for the number of neutrons observed at {\it MESSENGER}. This number of protons is at least an order of magnitude larger than that inferred from measurements of the largest gamma-ray line flares observed to date.  \citet{shar11} discussed this issue in more detail. 

The essential point that we are making is that the claimed detections of neutrons from the Sun on 2007 December 31 and 2011 June 4 are not substantiated by either the observations or the simulations made by \citet{feld10} and \citet{lawr14}.  The arguments they made in support of solar neutrons in both papers are complicated, tortuous, and in some cases erroneous.  There is a far simpler explanation for the observed transients because neutron production from ion interactions in spacecraft is ubiquitous, especially during SEP events. 

\begin{figure} 
\noindent\includegraphics[width=21pc]{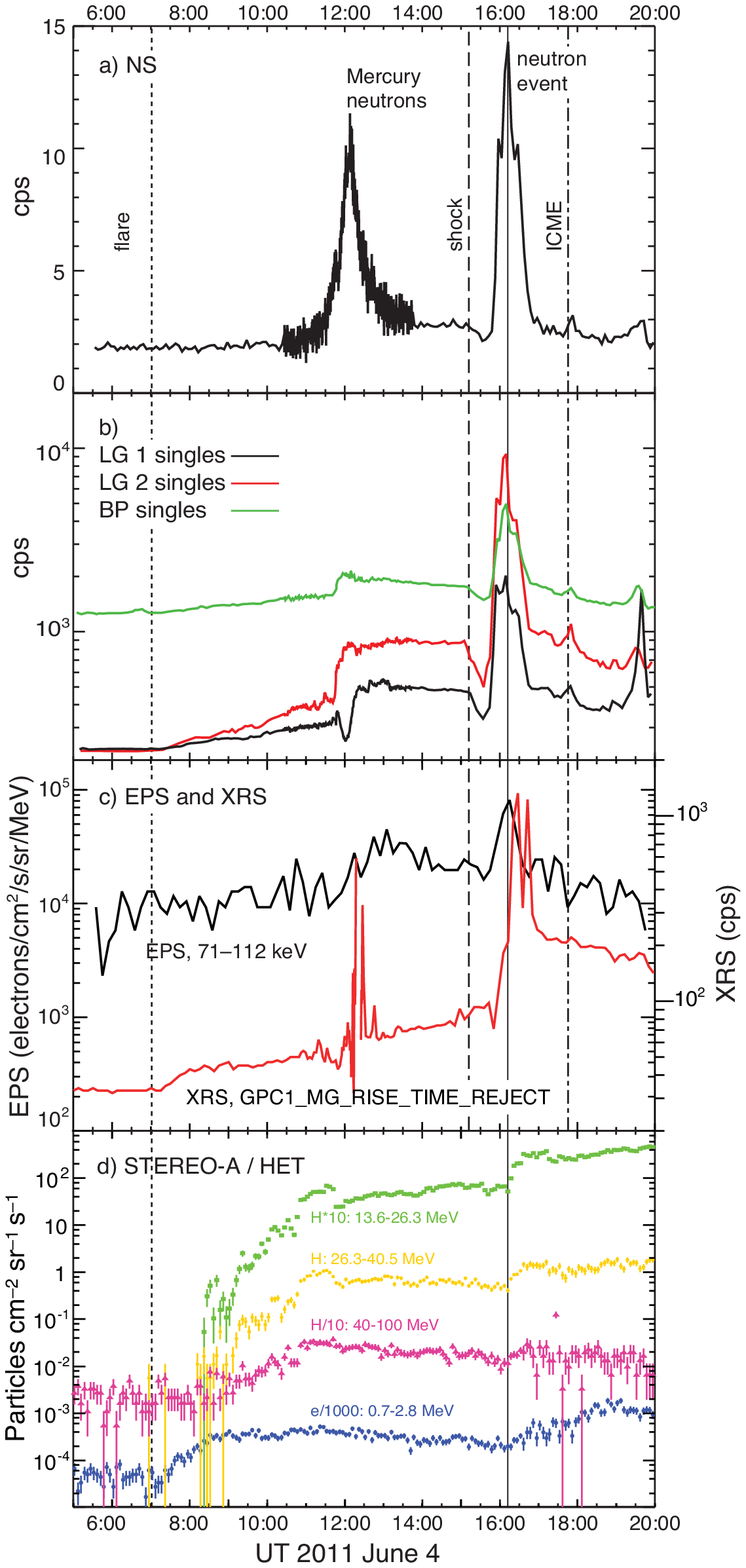}
%\includegraphics[width=20pc]{composite4_log.eps}
%\epsscale{.60}
%\plotone{composite3_linear.eps}
\caption{Time profiles of particle emissions observed on 2011 June 4 at {\it MESSENGER} and {\it STEREO A}.  a) {\it MESSENGER} NS valid neutron count rate; b) {\it MESSENGER} NS singles rates in the two lithium glass (LG1 and LG2) and borated plastic (BP) detectors; c) {\it MESSENGER} EPS 71--121 keV electron channel flux  and XRS particle rate from its rise-time reject channel; d) {\it STEREO A} IMPACT 5-min average fluxes of 0.7--2.8 MeV electrons and protons in various energy windows; scale factors have been applied for clarity.  Solid vertical line: peak of the neutron event; dotted line: onset of SEE at 07:06 UT; dashed line: start of passage of shock at {\it MESSENGER}; dot-dashed line: start of passage of CME at {\it MESSENGER}.}
\label{particles}
\end{figure}

\begin{figure}
\noindent\includegraphics[width=40pc]{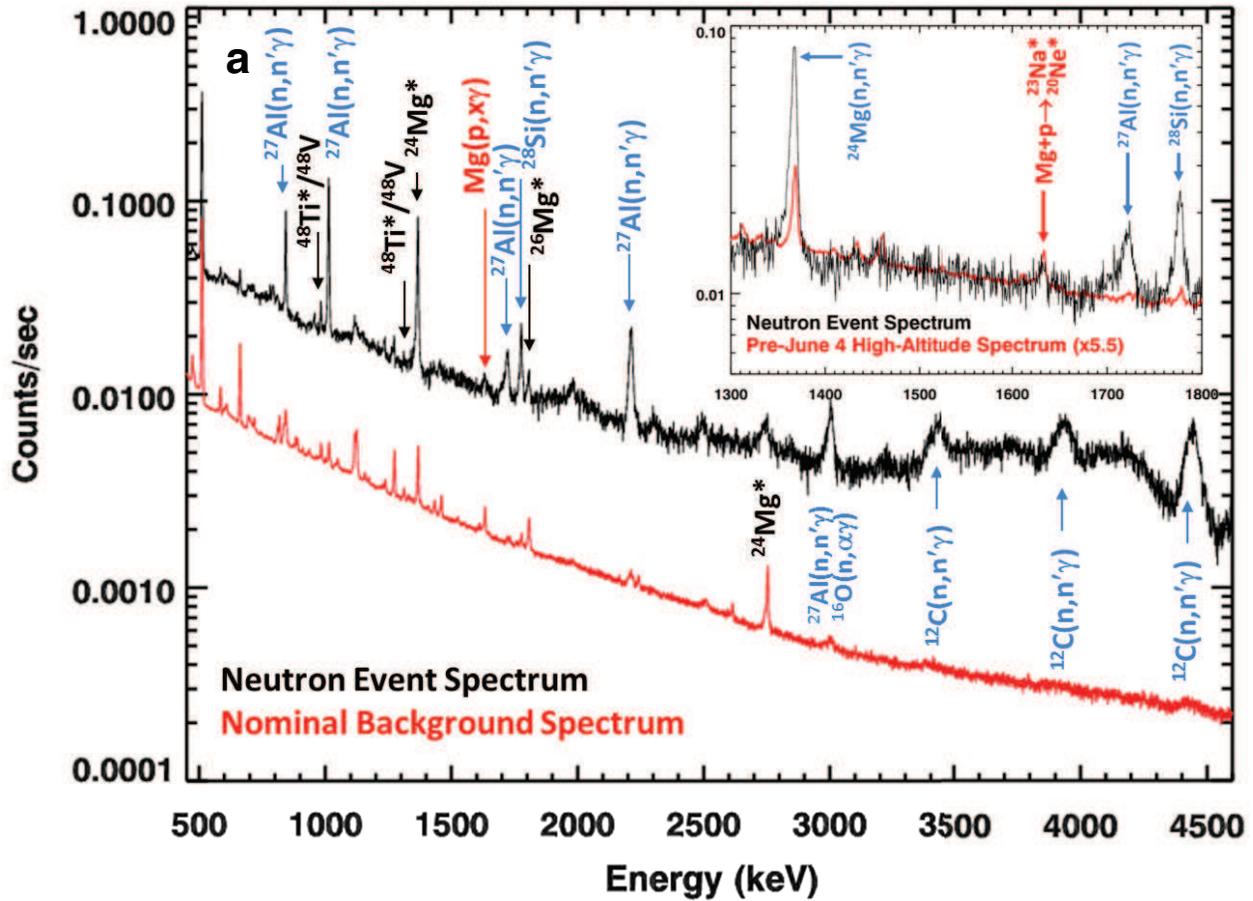}
%\plotone{fig2.eps}
\caption{{\it MESSENGER} GRS gamma-ray count spectra measured from 15:45 to 16:45 UT during the neutron transient (black) and during
background periods (red) reprinted from Figure 10 of \citet{lawr14}.  De-excitation lines produced from inelastic scattering and spallation (and their first and second positron escape peaks) are identified.  The authors incorrectly assert that only neutrons interacting in the spacecraft can produce the lines from inelastic scattering.  Protons interacting in the spacecraft produce the same lines with comparable cross sections.   The lines from spallation of Mg are also  identified as being produced only by proton interactions, but they too are produced by neutron interactions having comparable cross sections.  See text.}
\label{gamspec}
\end{figure}

Despite their failure to unambiguously detect solar neutrons in the inner heliosphere, we hope that the {\it MESSENGER} team will continue its search. We suggest that this search concentrate on times after high-energy SEEs with hard X-ray or gamma-ray emission.  The greatest chance for a successful and unambiguous detection will be for those events whose source on the Sun is not well connected magnetically with the location of {\it MESSENGER} so that the ions producing secondary neutrons at {\it MESSENGER} cannot reach the spacecraft until after the solar neutrons.  We do not expect detection of many events, because it is likely that only the largest gamma-ray flares would produce a detectable neutron signal in the NS.  From Figure \ref{particles}a we estimate that the increase in neutron rate would need to exceed about 0.5 counts s$^{-1}$ to be detectable with NS.  If the event lasted one hour, we estimate that at least 10$^{32}$ protons with energies $>$30 MeV having a power-law differential spectrum with an index of --4 would be required.  Such proton numbers are typically found in only the largest gamma-ray flares.   From experience with the 2007 December 31 and 2011 June 4 events, it is essential that the study only be done when there is no evidence for interplanetary protons at {\it MESSENGER}.

%%% End of body of article:

%%%%%%%%%%%%%%%%%%%%%%%%%%%%%%%%
%% Optional Appendix goes here
%
% \appendix resets counters and redefines section heads
% but doesn't print anything.
% After typing \appendix
%
%\section{Here Is Appendix Title}
% will show
% Appendix A: Here Is Appendix Title
%
%%%%%%%%%%%%%%%%%%%%%%%%%%%%%%%%%%%%%%%%%%%%%%%%%%%%%%%%%%%%%%%%
%
% Optional Glossary or Notation section, goes here
%
%%%%%%%%%%%%%%
% Glossary is only allowed in Reviews of Geophysics
% \section*{Glossary}
% \paragraph{Term}
% Term Definition here
%
%%%%%%%%%%%%%%
% Notation -- End each entry with a period.
% \begin{notation}
% Term & definition.\\
% Second term & second definition.\\
% \end{notation}
%%%%%%%%%%%%%%%%%%%%%%%%%%%%%%%%%%%%%%%%%%%%%%%%%%%%%%%%%%%%%%%%
%
%  ACKNOWLEDGMENTS

\begin{acknowledgments}
We thank Anne K. Tolbert for assistance in accessing and plotting {\it MESSENGER} XRS data and Richard Starr (richard.d.starr@nasa.gov) for explaining how to access to these data and the characteristics of the XRS instrument.  All of the other data used in this Comment either came from \citet{lawr14} or open sources such as the {\it STEREO} data center.  This work was funded in part by NSF/SHINE grant 1156092 and by the Office of Naval Research.  
\end{acknowledgments}

\end{article}
%
%
%% Enter Figures and Tables here:
%
% DO NOT USE \psfrag or \subfigure commands.
%
% Figure captions go below the figure.
% Table titles go above tables; all other caption information
%  should be placed in footnotes below the table.
%
%----------------
% EXAMPLE FIGURE
%
% \begin{figure}
% \noindent\includegraphics[width=20pc]{samplefigure.eps}
% \caption{Caption text here}
% \label{figure_label}
% \end{figure}

%
% ---------------
% EXAMPLE TABLE
%
%\begin{table}
%\caption{Time of the Transition Between Phase 1 and Phase 2\tablenotemark{a}}
%\centering
%\begin{tabular}{l c}
%\hline
% Run  & Time (min)  \\
%\hline
%  $l1$  & 260   \\
%  $l2$  & 300   \\
%  $l3$  & 340   \\
%  $h1$  & 270   \\
%  $h2$  & 250   \\
%  $h3$  & 380   \\
%  $r1$  & 370   \\
%  $r2$  & 390   \\
%\hline
%\end{tabular}
%\tablenotetext{a}{Footnote text here.}
%\end{table}

% See below for how to make sideways figures or tables.

\end{document}